\documentclass[lettersize,journal]{IEEEtran}
\PassOptionsToPackage{hidelinks}{hyperref}
\usepackage{hyperref}
\usepackage{color,array,amsthm}
\IEEEoverridecommandlockouts
\usepackage{cite}
\usepackage{amsmath,amssymb,amsfonts}
\usepackage{algorithm}
\usepackage{algorithmic}
\usepackage{graphicx}
\usepackage{textcomp}
\usepackage{xcolor}
\usepackage{makecell} 

\usepackage{enumitem} 
\graphicspath{ {images/} }

\begin{document}


\title{Optimized Contact Plan Design for Reflector and Phased Array Terminals in Cislunar Space Networks}

\author{
	Huan Yan,
	Juan A. Fraire,~\IEEEmembership{Member,~IEEE,}
	Ziqi Yang,
	Kanglian Zhao,~\IEEEmembership{Member,~IEEE,}
	Wenfeng Li,~\IEEEmembership{Member,~IEEE,}
	Jinjun Zheng,
	Chengbin Kang,
	Huichao Zhou,
	Xinuo Chang,
	Lu Wang,
	Linshan Xue,
	
	\thanks{H. Yan, Z. Yang, K. Zhao, W. Li, X. Hou, H. Li and Y. Miao are with Nanjing University, Nanjing 210008, China, E-mail: (yanhuan@smail.nju.edu.cn; 652022230035@smail.nju.edu.cn; zhaokanglian@nju.edu.cn; leewf\_cn@hotmail.com; houxiyun@nju.edu.cn; lihh@smail.nju.edu.cn; dz21260005@smail.nju.edu.cn); J. A. Fraire is with the CONICET, Instituto de Estudios Avanzados en Ingeniería y Tecnología (IDIT), Córdoba 5000, Argentina, and also with Saarland University, Saarbrücken 66123, Germany, E-mail: (juan.fraire@inria.fr); J. Zheng, C. Kang, H. Zhou, X. Chang, L. Wang, L. Xue are with the China Academy of Space Technology, Beijing 100094, China, E-mail: (zhjinjun@vip.sina.com; 75012565@qq.com; zhc202412@163.com; chang\_xinuo@foxmail.com; roadlu\_wang@qq.com; 201811040822@std.uestc.edu.cn). (Corresponding authors: Jinjun Zheng; Kanglian Zhao.)}
	
}



\maketitle

\begin{abstract}
Cislunar space is emerging as a critical domain for human exploration, requiring robust infrastructure to support spatial users-spacecraft with navigation and communication demands. 
Deploying satellites at Earth-Moon three-body orbits offers an effective solution to construct cislunar space infrastructure (CLSI). 
However, scheduling satellite links to serve users necessitates an appropriate contact plan design (CPD) scheme.
Existing CPD schemes focus solely on inter-satellite link scheduling, overlooking their role in providing services to users.
This paper introduces a CPD scheme that considers two classes of satellite transponders: Reflector Links (RL) for high-volume data transfers and Phased Array Links (PL) for fast switching and navigation services. 
Our approach supports both satellites and spatial users in cislunar space.
Simulations validate the scheme, demonstrating effective support for user while meeting satellite ranging and communication requirements.
These findings provide essential insights for developing future Cislunar Space Infrastructures.
\end{abstract}


\section{INTRODUCTION}
\label{sec_intro}

The increasing momentum of lunar and deep space exploration drives a surge in missions across cislunar space~\cite{1,2}. 
Consequently, spatial users—spacecraft running in cislunar space requiring communication and navigation support—face critical challenges. 
Due to the lower orbital altitude, these users experience degraded GNSS signal quality and poor ranging geometry~\cite{3,4}.
Furthermore, inherent limitations of ground stations (GSs), such as deployment constraints and Earth's rotation, prevent them from providing real-time service, as they are not continuously visible to users.
These challenges underscore the urgent need for a dedicated Cislunar Space Infrastructure (CLSI) to reliably support operations~\cite{33,5,6,7}.

In short, CLSI is an engineered system providing communication and navigation services for user in cislunar space.
Deploying satellites on cislunar three-body orbits provides a feasible approach to constructing CLSI. 
Key orbits include Earth-Moon libration point orbits (L1 to L5)~\cite{8,34}, Distant Retrograde Orbits (DROs)~\cite{56}, Elliptical Lunar Frozen Orbits (ELFOs)~\cite{55}, and Near-Rectilinear Halo Orbits (NRHOs)~\cite{54}.

In the cislunar space, certain users-such as situational awareness spacecraft—require high-volume data return to Earth.
To meet these high-capacity demands, satellites must support high-rate links.
Both laser terminals and large-aperture reflector antennas~\cite{22,50,51} provide high data rates.
However, due to inherent stability challenges with laser link pointing, acquisition, and tracking (PAT) over cislunar distances~\cite{57}, this work assumes satellites use more stable large-aperture reflector antennas (several meters in diameter).
These reflector terminals utilize parabolic reflectors to concentrate signals and rely on mechanical rotation for precise but relatively slow alignment, enabling reflector links (RLs) formation. 

\begin{figure}[] 
	\centering
	\includegraphics[width=\linewidth]{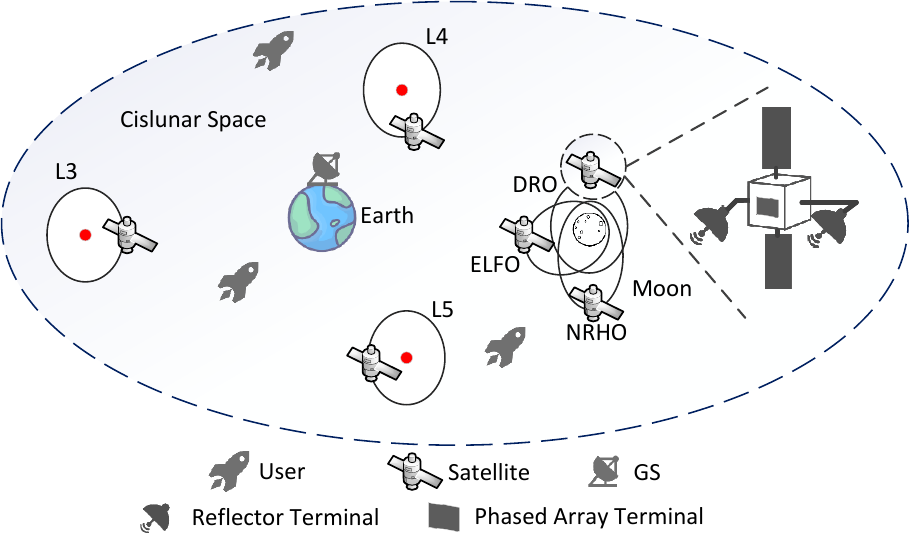}
	\caption{An Example of Cislunar Space Infrustructure }
	\label{fig1}
\end{figure}

For users requiring high-precision orbit determination, we assume each satellite is equipped with a time-division phased array terminal~\cite{23}-similar to GNSS satellites-to provide navigation services via phased array links (PLs).
These terminals consist of multiple small antennas that electronically steer beams by adjusting signal phase and amplitude, enabling PLs with rapid beam agility in direction and gain.
Additionally, satellites require inter-satellite ranging for improved orbit-determination accuracy and inter-satellite communication to relay telemetry to GS. PLs also meet these needs.
Fig.~\ref{fig1} illustrates an example CLSI, where satellites in cislunar three-body orbits carry both reflector antennas and a phased array antenna.

Determining how to schedule RLs and PLs to best satisfy the communication and navigation (ranging) needs of both users and satellites in cislunar space is a key challenge in Contact Plan Design (CPD)~\cite{14,15,16,45}.
Table~\ref{core difference between RL and PL} summarizes critical differences between RLs and PLs.
Existing CPD schemes exhibit the following limitations in the considered scenario:
\begin{table}[]
	\caption{Key Characteristics of RL and PL}
	\label{core difference between RL and PL}
	\centering
	\begin{tabular}{|c|c|c|c|}
		\hline
		& Date Rate & Swing Speed &Main Function \\
		\hline
		RL & High ($\approx$10 Mbps) & Low ($\approx$0.3 Deg/s) & Communication\\
		\hline
		PL & Low ($\approx$10 Kbps) & High (Immediate) & Navigation\\
		\hline
	\end{tabular}
	
\end{table}
\begin{enumerate}
	\item The reflector link topology (R-Topo) evolves slowly due to mechanical steering and is primarily used for high-volume data transfer. As such, RL scheduling shares similarities with prior CPD schemes for communication constellations~\cite{49,47,48,24}.
	However, existing schemes exclusively address inter-satellite link scheduling, neglecting satellite-user link planning.
	When RLs solely interconnect satellites, they form a densely connected network. This enables persistent end-to-end paths, allowing conventional CPD schemes to optimize metrics like minimizing inter-satellite delay or maximizing aggregate throughput.
	However, when satellites allocate RLs to serve users, the remaining inter-satellite RLs may form a sparsely connected, potentially partitioned topology.
	Consequently, user data must traverse the time-varying R-Topo over multiple scheduling periods to reach Earth, forming multi-hop, store-and-forward routes.
	This transforms the network into a classic delay/disruption tolerant network (DTN)~\cite{14,30}, shifting the optimization objective from throughput maximization to completing user data delivery within the fewest possible periods.
	\item 
	The Phased Array Link topology (P-Topo) evolves rapidly due to electronic beam steering and is primarily intended for ranging and low-volume data transfer, making its scheduling analogous to prior GNSS CPD~\cite{17,18,19,20,21,35,32,36,37,38,39}.
    However, conventional GNSS schemes focus exclusively on allocating PLs to meet satellites' own ranging and communication demands, disregarding user-directed PL assignments.
	Allocating PLs to users inevitably consumes resources that would otherwise support critical satellite operations. 
	Thus, the challenge is to deliver high-quality navigation and communication services to users while ensuring that satellites’ essential ranging and inter-satellite communication capabilities remain uncompromised.
	\item Moreover, prior CPD approaches typically allocate only one link type per scheme. Our work addresses this gap by co-scheduling RLs and PLs—requiring an integrated design that tightly couples R-Topo and P-Topo evolution.
\end{enumerate}

To address these gaps, we propose dedicated CPD schemes for RLs (R-CPD) and PLs (P-CPD).
To the best of our knowledge, this work is the first to design and solve an overall CPD framework that schedules RLs and PLs to simultaneously satisfy satellites’ ranging and communication requirements and the service demands of spatial users in a cislunar environment.
The key contributions are:
\begin{enumerate}
	\item \textit{Reflector-Terminal CPD (R-CPD)}: 
	We develop a CPD scheme tailored for reflector terminals within CLSI. 
	By formulating the problem as an integer linear program (ILP), we generate an superior R-Topo. 
	This topology achieves dual objectives: enabling high-volume user data delivery to GS in small scheduling periods while greatly enhancing P-Topo performance.
	\item \textit{Phased Array Terminal CPD (P-CPD)}: 
	Building on the R-Topo foundation, we design a tightly coupled CPD scheme for phased array terminals that maximizes R-Topo benefits. Using the Maximum Weight Matching Algorithm~\cite{25}, we construct P-Topo to simultaneously address satellite ranging/communication needs and provide user navigation/low-volume data services. 
	\item \textit{Extensive Simulations and Validation}: 
	We validate the proposed CPD schemes through comprehensive simulations following Reference~\cite{8}, deploying satellites at Earth-Moon L3, L4, L5 libration points and a lunar Distant Retrograde Orbit (DRO). 
	Results demonstrate that R-CPD achieves significantly lower user delay than LAA-PMM while its generated R-Topo markedly improves P-Topo performance over the same baseline.
	Furthermore, P-CPD outperforms DFCP across satellite communication, ranging accuracy, and user service metrics, simultaneously exhibiting superior adaptability to R-Topo dynamics.
	These results provide the first performance benchmark for CPD-based cislunar networks.
\end{enumerate}

The paper is organized as follows:
Section~\ref{sec_background} reviews related work on contact plan design and cislunar infrastructure.
Section~\ref{sec_model} details the network model of CISL, including its node architecture and link characteristics. 
Section~\ref{sec_rcpd} presents R-CPD, the ILP model and optimization strategies for the R-Topo.
Section~\ref{sec_pcpd}  analyzes how the R-Topo affects phased array terminals and proposes P-CPD for the P-Topo.
Section~\ref{sec_evaluation} evaluates the performance of the R-CPD and P-CPD topologies through simulations.
Finally, Section~\ref{sec_conclusion} concludes the paper with key findings.

\section{BACKGROUND}
\label{sec_background}

This section provides context and related research on Cislunar Space Networks and Contact Plan Design. 

\subsection{Cislunar Space Infrastructures}

\subsubsection{Actors and Projects}
The growing international collaboration highlights the critical role of cislunar infrastructure in enabling sustained lunar exploration, resource utilization, and the expansion of human presence beyond Earth.

\paragraph{LunaNet (NASA)}
NASA’s LunaNet~\cite{6} envisions a scalable lunar communication and navigation architecture that integrates topological configurations, including surface-based and orbiting provider nodes. 
This initiative aims to support the growing demand for communication and navigation services across manned and unmanned lunar missions.

\paragraph{Moonlight (ESA)}
Similarly, ESA’s Moonlight initiative~\cite{40} focuses on establishing the Lunar Communications and Navigation Services system and its associated infrastructure. 
The program proposes deploying a network of spacecraft around the Moon, enabling comprehensive support for lunar exploration activities while fostering collaboration with European aerospace companies to deploy communication and navigation satellites.

\paragraph{Lunar Navigation Satellite System (JAXA, NASA, and ESA)}
The Japan Aerospace Exploration Agency (JAXA) has introduced the Lunar Navigation Satellite System (LNSS)~\cite{41}, a constellation modeled after GNSS to provide navigation services and facilitate Earth-Moon communication relay. 
NASA, ESA, and JAXA are collaborating to establish the first batch of these lunar network service nodes. 
This initial network comprises two NASA Lunar Communication Relay Navigation System satellites, one ESA Moonlight Program satellite, and one JAXA LNSS satellite. 
All operate in circumlunar elliptical frozen orbits to provide continuous communication and navigation services, particularly focused on the Moon’s south pole.

\paragraph{China Aerospace Science and Technology Corporation (CASC)}
China Aerospace Science and Technology Corporation (CASC) has proposed a comprehensive Cislunar infrastructure~\cite{33}, comprising five key components: ground facilities, near-Earth constellations, near-lunar constellations, extended space constellations, and lunar surface facilities. 
The extended space constellations are of particular interest. 
These involve deploying satellites at the Earth-Moon libration points, providing uninterrupted, all-space services for users across the Cislunar domain.

\subsubsection{Onboard Terminals}

Adopting large-aperture reflector antennas on satellites is one of the solutions to meet the growing data demand in space systems~\cite{22}. The work in~\cite{52} provides the current progress status of large deployable spaceborne reflector antennas and reveals the unique advantages of using high-gain reflector antennas in space telecommunications, earth observation, and space science. 
Authors in~\cite{50} describe the design, ground testing, in-orbit experiments, and a in-orbit operation for large deployable reflector antennas. Two large deployable reflector antennas are installed on Engineering Test Satellite VIII for the experiment. 
~\cite{51} briefly introduces the TerreStar satellite: the TerreStar-1 is the world's largest commercial satellite, which communicates with tiny earth terminals and achieves very high antenna gain to receive uplink signals through a large-aperture reflector antenna.

However, mechanically scanned reflector antennas have large inertia and slow steering speeds. 
Adopting phased array antennas is one of the best approaches to overcoming these challenges. Phased array antennas are characterized by rapid beam scanning and agile beam shaping, among other features. 
The study in~\cite{23} summarizes the development process of spaceborne active phased array antennas, analyzing the structural forms, performance requirements, and application scenarios of spaceborne active phased array antennas. 
The BeiDou Navigation Satellite System employs a Ka-band phased array antenna to establish inter-satellite links that integrate both ranging and communication functions, thereby substantially improving the orbit determination accuracy of BeiDou satellites~\cite{53}.

\subsection{Contact Plan Design}

A contact is an opportunity to establish a temporary communication link between two nodes when physical requirements, such as antenna alignment and sufficient received power, are met. 
Such feasible contacts within a network during a specific interval constitute the contact topology. 
From this, a subset of selected contacts forms the contact plan, which determines the actual links to be implemented, considering constraints such as interference, power, and resource limitations. 
The process of determining this optimal subset is called Contact Plan Design (CPD)~\cite{14}.

\subsubsection{CPD for Navigation Systems}

For GNSS (Global Navigation Satellite Systems), the primary objectives of CPD include maximizing ranging opportunities, achieving frequent and diverse observations, and minimizing end-to-end data delivery delay. 
CPD research for GNSS can be broadly categorized based on the methodologies employed:

\begin{enumerate}
	\item \textit{Linear Programming:} Studies such as~\cite{17} optimize communication performance while adhering to GNSS-specific constraints, such as consistent ranging. Similarly,~\cite{19} extends polling mechanisms to ground-satellite links, aiming to minimize the average data delivery delay from satellites to ground stations.
	\item \textit{Specific Rule-Based Approaches:} The work in~\cite{36} introduces a grouping-based strategy for improving CPD efficiency, optimizing both ranging observations and low-delay communications. Authors in~\cite{21} outline a three-step link scheduling method that combines a genetic algorithm with prioritized downlink routes. Additionally,~\cite{37} enhances link allocation with an adaptive topology optimization algorithm that leverages prior knowledge for iterative improvement.
	\item \textit{Matching Algorithms:} A distributed CPD approach is first proposed in~\cite{18}, while~\cite{20} employs a hybrid of bipartite and general matching to improve inter-satellite ranging and communication performance. In~\cite{39}, links are classified into four categories with tailored weight adjustments to enhance CPD outcomes.
	\item \textit{Heuristic Algorithms:} The work in~\cite{32} formulates CPD as a constraint optimization problem and solves it using simulated annealing, while~\cite{35} combines genetic algorithms with the Blossom algorithm to achieve optimal matching for satellite sequences. In~\cite{38}, a double-loop optimization algorithm addresses complex scheduling challenges in CPD.
\end{enumerate}
\subsubsection{CPD for Communication Systems}

The CPD objective often shifts towards minimizing end-to-end transmission delay and maximizing throughput for communication constellations. 
Unlike GNSS, communication constellations frequently utilize relatively static topologies with satellites equipped with multiple inter-satellite link (ISL) terminals. 
Early constellations, such as Iridium~\cite{49}, adopt a fixed topology where each satellite establishes links with adjacent satellites in the same orbit and neighboring orbital planes. 
In~\cite{47}, a repeating “motif” pattern is introduced for ISLs, enhancing network capacity with slight sacrifices in delay. 
However, this approach avoids establishing links between satellites moving in opposite directions, limiting flexibility. 
Dynamic optimization of inter-plane links is addressed in~\cite{48}, which proposes inter-plane links to reduce delays in extensive low-Earth orbit constellations. 
Leveraging a perfect-matching model, \cite{24} generates a topology that simultaneously achieves 100 \% link utilization and a superior average inter-node distance.

Despite these advances, existing CPD schemes primarily optimize inter-satellite links while neglecting user service requirements. 
This paper bridges this gap by proposing a CPD framework for satellites in Earth-Moon three-body orbits equipped with both reflector and phased array terminals. 
Our approach simultaneously satisfies satellite ranging/communication needs and prioritizes user services through coordinated RL/PL scheduling. 
This integrated strategy robustly serves both system and user demands.

\section{SYSTEM MODEL}
\label{sec_model}
The CLSI addresses communication and navigation requirements for both satellites and users. 
The specific system model adopted in this work is detailed below.

\subsection{Terminal Configuration}
In this system, each satellite carries multiple reflector terminals and one phased array terminal.
Both RLs and PLs are bidirectional, supporting two-way communication and ranging.
Spatial users are equipped with either a reflector terminal (R-User) or a phased array terminal (P-User), enabling  access to satellites.
R-Users require high-volume data transmission, while P-Users need navigation or low-volume data services.
GSs can accommodate an arbitrary number of reflector terminals to facilitate high-capacity data links.
\subsection{Topology Model}

The CLSI topology model combines different operational schemes for phased array links (PLs) and reflector links (RLs) to match their distinct characteristics. 
PLs leverage Time Division Duplexing (TDD) for efficient and flexible link utilization, while RLs rely on a Finite State Automaton (FSA) scheme to minimize frequent link switching and maintain stability~\cite{26}.
In the FSA scheme, the network operates in discrete states with static visibility between nodes. 
Two nodes are visible in a state only if they maintain uninterrupted visible throughout that state.
This approach ensures that RL topology (R-Topo) remains stable during each FSA state, aligning with the mechanical constraints of RLs.

\begin{figure}[] 
	\centering
	\includegraphics[width=0.8\linewidth]{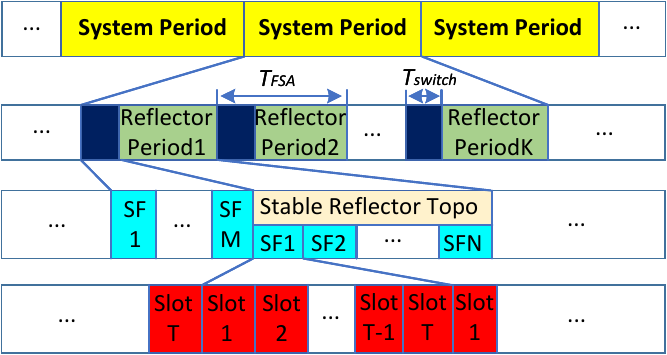}
	\caption{CLSI topology model}
	\label{fig2}
\end{figure}

To integrate TDD and FSA effectively within CLSI, the operational timeline is divided into equal-length reflector periods, as depicted in Fig.~\ref{fig2}. 
Each reflector period begins with a reserved interval dedicated to RL switching. 
During this interval, only non-switching-RLs and PLs are operational. 
Once this phase concludes, the R-Topo stabilizes for the remainder of the reflector period.

Reflector periods are further segmented into superframes, which serve as the scheduling units for phased array-based CPD. 
Each superframe consists of multiple time slots, representing the unit duration required to establish a PL. 
Within each slot, satellites utilizing PLs execute tasks such as ranging and data exchange~\cite{28}. 
This hierarchical time structure balances the stable but slow dynamics of RLs with the agility and fast switching of PLs, enabling seamless CLSI operation.

\subsection{Link Function}
\begin{figure}[] 
	\centering
	\includegraphics[width=\linewidth]{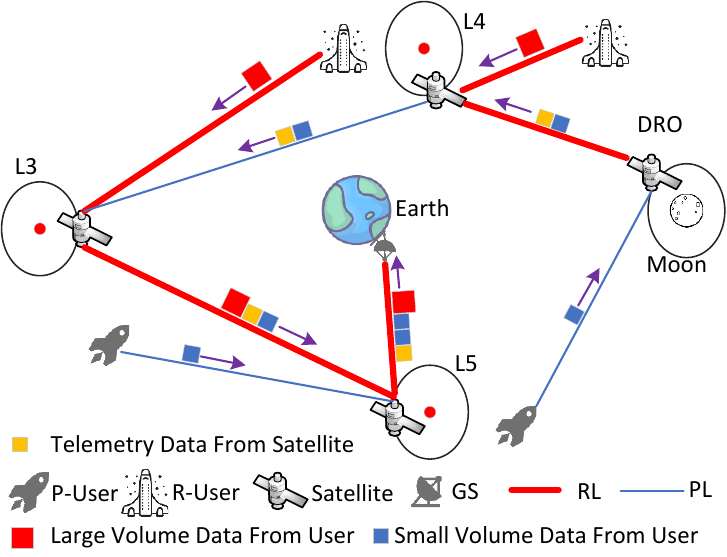}
	\caption{CLSI Link and Traffic Transmission Example}
	\label{fig3}
\end{figure}

The network supports various types of links, as illustrated in Fig.~\ref{fig3} and enumerated in Table~\ref{tab_links}: reflector links (RL) between satellites, phased array links (PL) between satellites, RLs between satellites and users, PLs between satellites and users, and RLs between satellites and ground stations (GS). 
Users in this system cannot establish direct links with one another or GSs; instead, all data must be relayed via satellites.

\begin{table}[]
	\centering
	\caption{Types of Links in CLSI}
	\label{tab_links}
	\begin{tabular}{|l|l|}
		\hline
		\textbf{Link Type} & \textbf{Description} \\ \hline
		RL(Sat,Sat) & Reflector links between satellites. \\ \hline
		PL(Sat,Sat) & Phased array links between satellites. \\ \hline
		RL(Sat,User) & Reflector links between satellites and users. \\ \hline
		PL(Sat,User) & Phased array links between satellites and users. \\ \hline
		RL(Sat,GS) & Reflector links between satellites and GSs. \\ \hline
	\end{tabular}
\end{table}

The system distinguishes the roles of these links based on their technical capabilities. 
About the user access, RL(Sat,User) are employed to deliver large-volume communication to R-Users, whereas PL(Sat,User) provide navigation and small-volume communucation to P-Users.
Within the satellite core network, both RL(Sat,Sat) and PL(Sat,Sat) are capable of inter-satellite ranging.
RL(Sat,Sat)s can relay both large- and small-volume inter-satellite data, while PL(Sat,Sat)s are restricted to small-volume inter-satellite data forwarding.
For satellite-to-GS connectivity, RL(Sat,GS)s are used to establish high-bandwidth links.

Users are assumed to have only access capabilities, meaning they transmit their data to a connected satellite as soon as a link is established. 
Satellites then compute end-to-end paths using a routing algorithm~\cite{29} within a store-and-forward network, adhering to the principles of Delay/Disruption Tolerant Networking (DTN)~\cite{14,30}. 

RLs and PLs jointly work to fulfill the system’s communication and navigation functions. 
Large-volume data from R-users is exclusively transmitted through the evolving R-Topo over multiple reflector periods. 
In contrast, small-volume data generated by P-users or satellites is delivered through the evolving P-Topo across multiple time slots, augmented by the R-Topo within the current reflector period. 
Fig.~\ref{fig3} illustrates a traffic delivery process within CLSI.

This collaborative framework ensures efficient management of diverse traffic requirements within the CLSI.

\subsection{Traffic Consideration}

The primary traffic in CLSI comprises data transmitted to ground stations (to-GS traffic), such as large-volume observational data from R-Users and the telemetry data from P-Users and satellites.    
Accordingly, the subsequent CPD schemes are designed to minimize to-GS delay for this traffic.
Secondary traffic, including inter-satellite navigation messages, is not considered in the optimization objective.

\section{R-CPD METHOD}
\label{sec_rcpd}

This chapter presents an ILP model for generating the R-Topo.
We use $V_s$, $V_u$, and $V_g$ to denote the sets of satellites, R-users, and GSs, respectively. 
The entire set of nodes is represented as $V=V_s \cup V_u \cup V_g$.
The index m  $\in \{1, 2, 3, ..., M\}$ represents the current reflector period.
A binary matrix $Y$ represents the visibility between nodes, where $y(i, j, m)$ = 1 indicates that node i and node j are visible to each other during the m-th reflector period, and 0 otherwise. 
Similarly, a binary matrix $X$ represents the link relationships between nodes, where $x(i, j, m)$ = 1 denotes that a link is established between node i and node j in the m-th reflector period, and 0 otherwise.

\subsection{Basic constraints}
The following are three general constraints concerning link relationships matrices $X$ and visibility matrices $Y$:
\begin{equation}
	x_{i,j,m}=\{0,1\}, \forall v_i,v_j \in V, m\in M,\label{eq1}
\end{equation}
\begin{equation}
	x_{i,j,m}=x_{j,i,m}, \forall v_i,v_j \in V,  m\in M,\label{eq2}
\end{equation}
and 
\begin{equation}
	x_{i,j,m}\leq y_{i,j,m}, \forall v_i,v_j \in V,  m\in M,\label{eq3}
\end{equation}
where \eqref{eq3} indicates that link establishment requires visibility.

Given that each satellite is equipped with $r$ reflector terminals and user carries 1 , we have:
\begin{equation}
	\sum_{v_i\in V} x_{i,j,m} \leq r,  \forall v_j \in V_s,  m\in M,\label{eq4}
\end{equation}
\begin{equation}
	\begin{split}
		\sum_{v_i\in V} x_{i,j,m} \leq 1,  \forall v_j \in V_u,  m\in M, \label{eq5}
	\end{split}
\end{equation}
We assume that the GSs have sufficient capacity, imposing no restriction on the number of links they can establish.
\subsection{R-User access constraints}
Given users’ inherent storage capability and the scarcity of satellite RL resources, access is not guaranteed in every reflector period; 
Instead, a minimum access frequency requirement is enforced for each user through a dedicated constraint:
\begin{equation}
	\ C_{i,m} =\{0,1\}, \forall v_i \in  V_u, m\in M.\label{eq6}
\end{equation}
\begin{equation}
	\ C_{i,m}=\sum_{v_j \in V_s}x_{i,j,k}, \forall v_i \in  V_u, m\in M.\label{eq7}
\end{equation}
\begin{equation}
	\ \sum_{m}^{m+f-1}C_{i,m}\geq 1, \forall v_i \in  V_u, 1 \leq m \leq M-f+1.\label{eq8}
\end{equation}
where $C_{i,m}$ indicates access occurrence for user i in period $m$, and \eqref{eq8} enforces a minimum access frequency $f$ per user.

\subsection{Satellite-GS connection constraints}
We set $L_G$ as the desired number of RL(Sat,GS). However, since the number of satellites simultaneously visible to the GS may fall below $L_G$, we introduce a slack variable $p(m)$ to represent the deficit:
\begin{equation}
p(m) \in \mathbb{N} \label{eq9}
\end{equation}

\begin{equation}
\sum_{v_i\in V_s, v_j \in V_g} x_{i,j,m} \geq L_G-p(m),  \forall m\in M, \label{eq10}
\end{equation}

\subsection{Maximize Inter-Satellite Connectivity Objective}
Subject to guaranteeing each user's prescribed access frequency and a minimum RL(Sat, GS)s availability, we aim to maximize inter-satellite connectivity; 
this supports both R-user data delivery to GS and subsequent P-CPD.
\begin{equation}
	\begin{split}
		max \sum_{v_i \in V_s, v_j \in V_s, m \in M} x_{i,j,m} - P* \sum_{m \in M}p(m)
		\\s.t. \eqref{eq1}-\eqref{eq10}.
		\label{eq11}
	\end{split}
\end{equation}
$P$ is a a sufficiently large constant; 
the objective value decreases by $P*\sum_{m \in M}p(m)$ for each unsatisfied RL(Sat,GS).
Consequently, apart from visibility constraints, the system endeavors to establish exactly $L_G$ RL(Sat,GS).

\section{P-CPD METHOD}
\label{sec_pcpd}

Each satellite in CLSI is equipped with a single phased array terminal~\cite{23}. 
The P-Topo is formed as pairs of nodes and can be planned using matching algorithms from graph theory~\cite{25}. 
In an undirected graph, matching represents a set of edges where no two edges share a common vertex, and the maximum weight matching corresponds to the matching with the largest sum of edge weights~\cite{31}. 
Building upon the R-Topo, this section proposes a heuristic algorithm for P-CPD using the maximum weight matching.

\subsection{The Impact of R-Topo on P-CPD}

\begin{figure}[] 
	\centering
	\includegraphics[width=\linewidth]{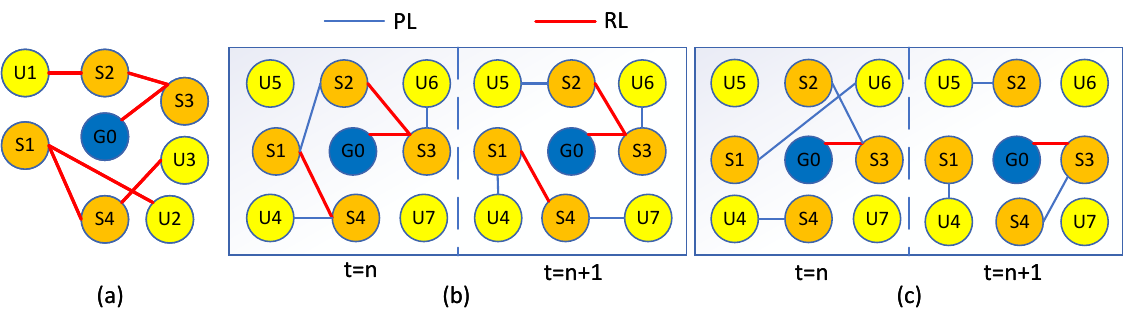}
	\caption{An example of the impact of R-Topo on P-CPD}
	\label{fig6}
\end{figure}

Fig.~\ref{fig6}(a) illustrates a sample R-Topo, where U, G, and S denote users, GSs, and satellites, respectively. Satellites directly (e.g., S3) or indirectly (e.g., S2 through RL(S2,S3)) connected to a GS are referred as G-Sats;
their set is denoted by $V^{gs}$. 
Satellites S1 and S4, with no GS connection, are termed UG-Sats.
P-CPD introduces three key modifications relative to the R-Topo foundation:
\begin{enumerate}
	\item Since RL inherently support ranging, satellites connected via RLs are exempt from additional PL-based ranging.
	\item For satellites indirectly connected to the GS (e.g., S2 in Fig.~\ref{fig6}(a)), data can be relayed to GS without PLs. Notably, the data considered in P-CPD pertains exclusively to small-volume data. 
	\item  If RL(UG-Sat, UG-Sat)s exist, all connected UG-Sats form a set denoted as $V_n^{ugs}$, where $n \in \mathbb{N}^+$ indexes the n-th such set.
	An isolated UG-Sat forms its own set. 
	when any UG-Sat within a UG-Sats set establishes a PL with a G-Sat, all UG-Sats in that set can relay data to GSs.
\end{enumerate}

Fig.~\ref{fig6}(b) demonstrates the P-CPD results over two-time slots based on R-Topo (excluding RL(Sat, User)s). 
At the $t=n$ slot, S1 established a PL with S2, enabling all UG-Sats in the UG-Sats set $V^{ugs}_1=[S1, S4]$ to relay data to GSs during this time slot. 
Across two time slots, P-CPD successfully transmitted data to GSs for all satellites and served six phased array users.
Moreover, RL(S1,S4) and RL(S2,S3) additionally provide ranging benefits.
In contrast, Fig.~\ref{fig6}(c) illustrates the P-CPD results over two-time slots without considering R-Topo (only retaining RL(Sat, GS)s as satellite-to-GS link). 
Here, P-CPD only succeeded in delivering data to GSs for S2, S3, and S4, while serving just four users. 

A comparison between Fig.~\ref{fig6}(b) and Fig.~\ref{fig6}(c) highlights key benefits brought by R-Topo to P-CPD:
The R-Topo enhances inter-satellite connectivity, enabling satellites to conserve PL resources otherwise used for ranging and communication; these resources can be reallocated to serve additional users or remain idle to reduce system energy consumption.
Correspondingly, the optimization objective of R-CPD is to maximize the inter-satellite connectivity.

\subsection{Mathematic Model and Algorithm}

\subsubsection{Basic Model}

We model the visibility between nodes' phased array terminals in each reflector period as an undirected graph $G(V,E,W)$, where $V$ is the set of all satellites and users. 
$E$ is the set of visible edges between all nodes' phased array terminals, i.e., the set of possible PLs. 
$W$ is the weight set of all visible edges. 
The visible edges between satellites and users, satellites and satellites, UG-Sats and UG-Sats, G-Sats and G-Sats, UG-Sats and G-Sats is denoted by $\varepsilon^{s-u}$,$\varepsilon^{s-s}$,$\varepsilon^{ugs-ugs}$,$\varepsilon^{gs-gs}$,$\varepsilon^{ugs-gs}$ respectively.

\subsubsection{Weights Assignment for Edges}
Table~\ref{parameters explanation} details the parameters used in the weight assignment process.

\paragraph{Edge Weight Between Satellites and P-Users}
$U_P$=$[(x,L^u_x),\dots,(z,L^u_z)]$ specifies each P-User's required PL count per superframe, where $(x,L^u_x)$ indicates user x requires $L^u_x$ PLs from satellites.
To satisfy $U_P$, define the weight $w_{i,j,t}^u$ of the $e_{i,j}$ between any user i and any satellite j in $G(V,E,W)$ at time slot t as follows:

\begin{equation*}
	w_{i, j, t}^{u}=\left\{ \begin{matrix} I_u(i,t)\cdot(L_{i}^{u}-L_{i,s,t})\cdot{C}_{u}, &C1~or~C2\\ 1, &C3 \\ 0 & C4\\ \end{matrix} \right. 
\end{equation*}
\begin{equation*}
C1: L_{i,s,t}<L_{i}^{u}~and~L_{i,s,t}<\bar{G(i)}~and ~L_{i,j,t}=0
\end{equation*} 
\begin{equation*}
C2: L_{i,s,t}<L_{i}^{u}~and ~L_{i,s,t} \geq \bar{G(i)}~~~~~~~~~~~~~~~~~~
\end{equation*}
\begin{equation*}
~C3:L_{i,s,t}<L_{i}^{u}~and~L_{i,s,t}<\bar{G(i)}~and ~L_{i,j,t}>0
\end{equation*}
\begin{equation*}
C4:L_{i,s,t}\geq L_{i}^{u}~~~~~~~~~~~~~~~~~~~~~~~~~~~~~~~~~~~~~~~
\end{equation*}
\begin{equation}
~~~~~~~~~~~~~~~~~~~~~~~~~~~~~~~~~~~~\forall e_{i, j} \in \varepsilon^{u-s}, 1 \leq t \leq T\label{eq12}
\end{equation}

\begin{equation*}
	I_u(i,t)=\left\{ \begin{matrix} 1, &t=1\\
		1,&\begin{aligned}
			&\exists e_{i,j}\in \varepsilon^{u-s} \cap M_{t-1}
		\end{aligned}  \\ 
		I_u(i,t-1)+1, &others  \end{matrix} \right. 
\end{equation*}
\begin{equation}
	~~~~~~~~~~~~~~~~~~~~~~~~~~~~~~~~~~~~1 \leq t \leq T\label{eq13}
\end{equation}
	\begin{table}[]
	\caption{Parameters Explanation}
	\label{parameters explanation}
	\centering
	\footnotesize
	\renewcommand{\arraystretch}{1.2}
	\begin{tabular}{|c|c|}
		\hline
		\textbf{Parameters} & \textbf{Explanation} \\
		\hline
		$T$ & the number of time slots in a superframe\\
		\hline
		$L_{i,j,t}$ & \makecell{the cumulative number of PLs between \\node i and node j up to the end of t-1 slot} \\ 
		\hline
		$L_i^{u}$ & the number of required PLs for user i \\
		\hline
		$L_{i,s,t}$ & \makecell{the total number of PLs all satellites \\provide to user i to the end of t-1 slot}\\
		\hline
	    $\bar{G(i)}$ & \makecell{the number of distinct satellites \\visible to the node i in graph G}  \\
		\hline
		$M_t$ & the P-CPD result in the t-th slot  \\
		\hline
		$C_u$ & service constant \\
		\hline
		$C_c$ & communication constant \\
		\hline
	    $C_r$ & ranging constant \\
		\hline
		$I_u(i,t)$ & access tendency for user i in t-th slot \\
		\hline
		$I_c(i,t)$  & ground tendency for satellite i in t-th slot \\
		\hline
	\end{tabular}
\end{table}
Essentially, $\bar{G(i)}$ is the maximum number of non-repeating satellites user i can link with.
The greater the diversity of a user’s link partners, the more diversified the satellite–user geometries during ranging, and hence the better the achieved navigation performance.
In~\eqref{eq12}:
\begin{enumerate}
	\item $C1$ indicates PL(i,j) will simultaneously contributes to satisfying user i’s required number of PLs and to increasing the diversity of its link partners.
    \item $C2$ denotes that user i has already attained optimal link partner diversity, so only the required number of PLs needs to be met.
    \item $C3$ signifies that user i’s link partner diversity remains sub-optimal; the  PL(i,j) can only increase the link count without improving partner diversity.
    \item $C4$ indicates that the required number of PLs for user i has been fully satisfied.
\end{enumerate}
The incorporation of $C1–C3$ promotes each user to establish links with a larger number of distinct satellites

When a user experiences prolonged satellite inaccessibility, its access tendency $I_u(i,t)$ grows; once the user reconnects, the tendency is reset to one. This mechanism enforces an approximately uniform PL(Sat,User) distribution within multiple slots, reducing average to-GS delay. 
Consequently, the weight assignment strategy between user and satellite jointly enhances both the communication and navigation services experienced by the user.
\paragraph{Edge Weight Between Satellites}
To optimize to-GS delay and satisfy inter-satellite ranging requirements, the edge weight between any satellite i and other satellite j in $G_P(V,E,W)$ is defined as follows at time slot t:
\begin{equation}
	w_{i, j, t}=\begin{matrix} w^c_{i,j,t}+w^r_{i,j,t}, &\forall e_{i, j} \in \varepsilon^{s-s}, 1 \leq t \leq T \end{matrix}  \label{eq14}
\end{equation}

Here, $w^c_{i,j,t}$ reflects the contribution of planning this PL to communication performance, While $w^r_{i,j,t}$ reflects the contribution of planning this PL to ranging performance.

In specific, the first part of ~\eqref{eq14} , $w^c_{i,j,t}$, can be given as:
\begin{equation*}
	w_{i, j, t}^c=\left\{ \begin{matrix} I_c(i,t)*C_c, &i \in V_n^{ugs},j \in V^{gs},if~i= S(V_n^{ugs})\\ 0, &i \in V_n^{ugs},j \in V^{gs},if~i \neq S(V_n^{ugs}) \\0,&\forall e_{i,j} \in \varepsilon^{gs-gs}\\0, & \forall e_{i,j} \in \varepsilon^{ugs-ugs} \end{matrix} \right. 
\end{equation*}
\begin{equation}
	~~~~~~~~~~~~~~~~~~~~~~~~~~~~~~~~~~1 \leq t \leq T\label{eq15}
\end{equation}

where $I_c(i,t)$  represents the grounding tendency of the UG-Sat i at the t-th time slot:
\begin{equation*}
	I_c(i,t)=\left\{ \begin{matrix} 1, &t=1\\
		1,&\begin{aligned}
			&((\exists e_{x,y}\in \varepsilon^{ugs-gs} \cap M_{t-1}) \\
			&\quad and~(x,i \in V_n^{ugs})) 
		\end{aligned}  \\ 
		I_c(i,t-1)+1, &others  \end{matrix} \right. 
\end{equation*}
\begin{equation}
	~~~~~~~~~~~~~~~~~~~~~~~~~~~~~~~~~~1 \leq t \leq T\label{eq16}
\end{equation}

~\eqref{eq16} signifies that for an UG-Sat i, its grounding tendency is initialized to 1 in the first time slot. If, in $M_{t-1}$, there exists a UG-Sat x in the UG-Sats set $V_n^{ugs}$  which i belongs to that has established a PL with any G-Sat y, then the grounding tendency of all UG-Sats in $V_n^{ugs}$, including UG-Sat i’s, are set to 1. 
Otherwise, UG-Sat i’s grounding tendency equals the grounding tendency of the previous time slot plus one. 
If set $V_n^{ugs}$ does not establish any PL with G-Sats across several consecutive time slots. The grounding tendency of all UG-Sats in this set will increase, thereby providing more opportunities for these UG-Sats to be linked with G-Sats in subsequent matches.

In~\eqref{eq15}, $S(V_n^{ugs})$ represents a randomly selected UG-Sats from $V_n^{ugs}$. 
We only allocate communication weight to the edges between the selected UG-Sat and G-Sats; 
For other UG-Sats in $V_n^{ugs}$ that have not been selected, we do not assign communication weight. 
This method allows $V_n^{ugs}$ to transmit data to GSs by establishing just one PL with G-Sats.
Any edge between UG-Sats or between G-Sats carries zero communication weight because such edges do not contribute to communication.

The second part of ~\eqref{eq14}, $w^r_{i,j,t}$, can be expressed as:
\begin{equation*}
	w_{i, j, t}^{r}=\left\{ \begin{matrix} 
		C_r, & L_{i,j,t}=0~and~RL(i,j)~inexistence\\ 
		0,  &L_{i,j,t}> 0~or~RL(i,j)~existence
		&\\ \end{matrix} \right. 
\end{equation*}
\begin{equation}
	~~~~~~~~~~~~~~~~~~~~~~~~~~~~~~~~~~\forall e_{i,j}\in \varepsilon^{s-s} , 1 \leq t \leq T\label{eq17}
\end{equation}

If $L_{i,j,t}>$0, meaning prior PL(i,j) establishment, repeated link establishment offers no significant benefit to ranging. 
As RL(i,j) inherently performs ranging, no additional ranging weight is required when RL(i,j) is active.
The proposed weight assignment strategy about $w_{i, j, t}^{r}$ and $w_{i, j, t}^{c}$ can fully exploit the communication and ranging benefits from the R-Topo.

\subsubsection{A Heuristic Based on Maximum Weight Matching}

Algorithm~\ref{alg2} illustrates the process of P-CPD based on maximum weight matching. For the k-th reflector period, let $H(V,E)$=$(M_1,M_2,\dots,M_T)$ denote the P-CPD result in a superframe over T time slots.

\begin{algorithm} 
	\caption{P-CPD based on maximum weight matching}
	\label{alg2}
	\begin{algorithmic}[1]
		\REQUIRE Constellation parameters, number of time slots T, the R-Topo, $C_u$, $C_c$, $C_r$, $U_P$
		\ENSURE Topology result $H(V,E)$ 
		\STATE \textbf{Begin}
		\STATE t=1
		\STATE Generate the graph $G(V,E,W)$
		\WHILE {$t \leq T$ }
		\STATE Set weight for edges in $G(V,E,W)$ according to \eqref{eq12}-\eqref{eq17}
		\STATE Compute the maximum matching 
		\STATE According to the matching result $M_t$, obtain the scheduled topology in the t-th time slot
		\STATE t=t+1
		\ENDWHILE 
		\STATE Construct $H(V,E)$ from $(M_1,M_2,\dots,M_T)$
		\STATE \textbf{End}
	\end{algorithmic}
\end{algorithm}

\section{EVALUATION}
\label{sec_evaluation}

This section evaluates R-CPD and P-CPD performance under varying parameters. 

Since the large volume traffic from R-Users may traverse multiple reflector periods-evolving R-Topo before reaching GS, we define the delay in R-Topo as the number of reflector periods required to backhaul the traffic to the GS.
Similarly, the small volume traffic from satellites and P-Users may traverse multiple slots-evolving P-Topo before reaching GS. Consequently, in P-Topo, delay is defined as the number of slots required to backhaul the traffic to the GS.
Given that the durations of reflector periods and time slots significantly exceed data propagation (typically in the order of 1 second in cislunar space) and transmission delays (typically in the order of milliseconds), this work excludes the latter two types of delays from consideration.

Our simulations employ the constellation in Fig.~\ref{fig3} with satellites at L3, L4, L5 and lunar DRO ~\cite{8}. This configuration offers three advantages:
\begin{enumerate}
	\item It provides complete coverage of the entire cislunar space, ensuring services for users at any location within the cislunar space.
	\item The constellation leverage GSs in Kashi, Jiamusi, and Sanya to ensure continuous visibility with at least one GS at all times.
	\item Reference~\cite{8} quantifies orbit determination accuracy and convergence time for this constellation. 
	As our work focuses exclusively on link-scheduling strategies, the orbit-determination results under different link policies provided in~\cite{8} validate the soundness of the proposed scheduling strategy.
\end{enumerate}

Users are evenly distributed across Earth-Moon libration points L1–L5, ELFO, NRHO, and lunar DRO.
Table~\ref{Scenario Parameters} lists GS locations, satellite orbits, and user trajectories adopted in the scenario.

Simulation spans 30 days, approximately one regression period of the constellation. 
Each reflector period comprises 12 superframes, with the first two superframe corresponding to the RLs switching time and the remaining ten superframes representing its stable phase.
Owing to the large aperture of the reflector terminals, we assume that each satellite carries two such antennas. 
Further detailed simulation parameters are summarized in Table~\ref{scene_table}.
\begin{table}[]
	\caption{Scenario Parameters}
	\label{Scenario Parameters}
	\centering
	\begin{tabular}{|c|c|}
		\hline 
		Satellite orbit  & L3,L4,L5,DRO\\
		\hline
		User orbit & L1, L2, L3, L4, L5, DRO, NRHO, ELFO\\
		\hline
		Jiamusi & (46.8°N, 130.3°E) \\
		\hline
		Kashi & (39.47°N, 75.99°E) \\
		\hline
		SanYa & (18.23°N, 109.02°E) \\
		\hline
		
	\end{tabular}
\end{table}
\begin{table}[]
	\caption{Basic parameters in the simulation}
	\label{scene_table}
	\centering
	\begin{tabular}{|c|c|}
		\hline
		Simulation Duration & 30 days \\
		\hline
		Length of a Reflector Period & 60 min \\
		\hline
		\makecell{The switching time of \\the reflector terminals} & 10 min \\
		\hline
		Length of a Superframe & 5 min \\
		\hline
		Length of a Time Slot & 10 s \\
		\hline
		RL Pointing Range & 75° \\
		\hline
		PL Pointing Range & 75° \\
		\hline
		GS Pointing Range & 85° \\
		\hline
		\makecell{Number of Reflector Terminals \\on the Satellite} & 2\\
		\hline
	\end{tabular}
\end{table}

\subsection{Performance of R-Topo}
We select the LAA-PMM~\cite{24} algorithm as the benchmark for R-CPD. 
LAA-PMM leverages a perfect-matching model to yield a communication topology with extremely high link utilization, thereby outperforming both MSN~\cite{58} and greedy~\cite{59} schemes in terminal utilization. 
The rationale for adopting LAA-PMM is that high terminal utilization implies enhanced overall connectivity, aligning with the inter-satellite connection optimization objective of the proposed R-CPD.
In R-CPD, the user access frequency is set to $f=2$, and the penalty constant in the objective function is set to $P=1000$ (all subsequent simulations employ these defaults).

\begin{figure}[tbp] 
	\centering
	\includegraphics[width=0.9\linewidth]{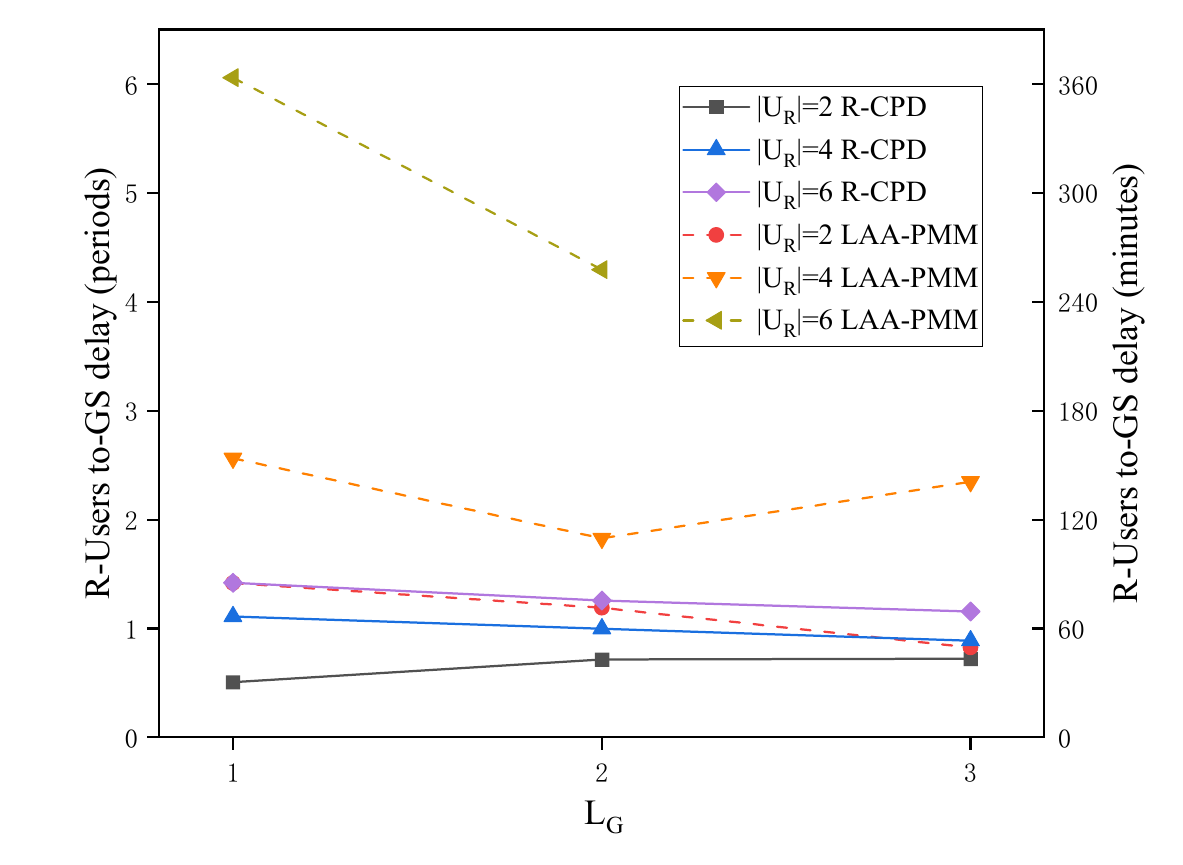}
	\caption{R-Users to-GS delay}
	\label{fig7}
\end{figure}

Fig.~\ref{fig7} compares average R-user to-GS delay between R-CPD and LAA-PMM across varying ground link counts ($L_G$) and R-user populations ($|U_R|$). 
R-CPD consistently achieves lower delay than LAA-PMM under identical conditions, maintaining stable delays of approximately one reflector period. 
This ensures bulk traffic from any R-User can be delivered to GSs within approximately one reflector period on average.
Conversely, LAA-PMM delays increase significantly with user population growth due to its terminal utilization maximization strategy.
LAA-PMM allocates a large number of satellite terminals to users in every period, thereby sacrificing inter-satellite connectivity.
This highlights the inadequacy of conventional communication constellation CPDs for such user access scenarios.
In contrast, R-CPD intermittently serves users, leaving abundant terminals available for RL(Sat,Sat)s; the resulting enhanced inter-satellite connectivity, in turn, facilitates efficient traffic backhaul.

\subsection{Performance of P-Topo}
Most existing GNSS CPD schemes (rule-based and ILP-based) require explicit node type classification (e.g.,  anchor/non-anchor satellite), failing when user nodes are introduced.  
In contrast, the matching-based GNSS CPD scheme-DFCP~\cite{18} schedules PL(User,Sat)s without node-type distinctions.
It dynamically adjusts edge weights and performs maximum-weight matching to maintain both satellite ranging and communication performance.
We benchmark P-CPD against DFCP since both use maximum-weight matching, enabling direct comparison of edge-weighting strategies. 
For fairness, DFCP is modified to cease PL(User,Sat) allocation once a user's link requirement is satisfied.
In all subsequent P-CPD, the service constant $C_u$, communication constant $C_c$ and ranging constant $C_r$ are fixed at $C_u=1$, $C_c=8$ and $C_r=30$. 
In addition, each P-user is assumed to require four links per superframe, i.e., $\forall (x,L_x^u) \in U_P$, $L_x^u$=4.

\subsubsection{Satellite Communication}
G-Sats are directly or indirectly connected to GSs, incurring a delay of zero slots. 
Therefore, only the to-GS delay of UG-Sats is considered.

\begin{figure}[tbp] 
	\centering
	\includegraphics[width=0.9\linewidth]{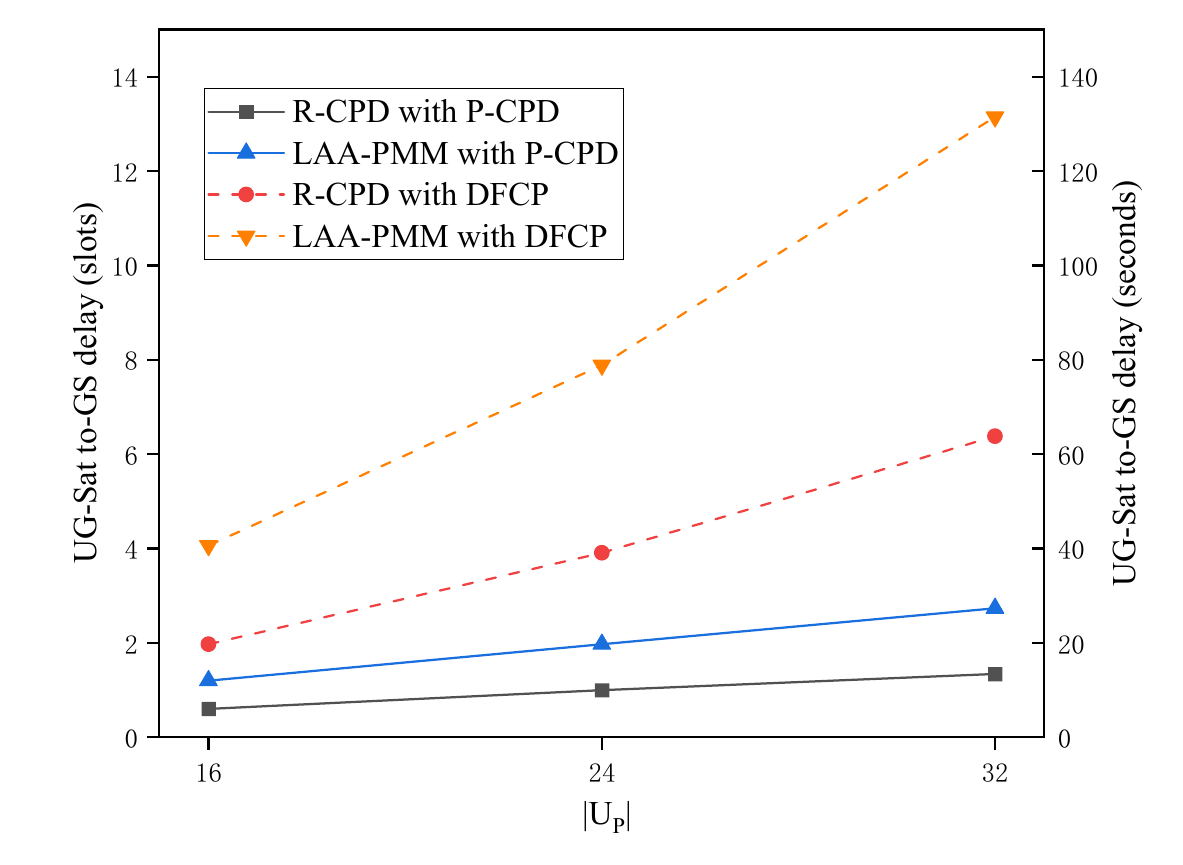}
	\caption{UG-Sats to-GS delay vs. $U_P$}
	\label{fig8}
\end{figure}

Fig.~\ref{fig8} shows UG-Sats to-GS delay under R-Topo driven by R-CPD or LAA-PMM, for $|U_R|$=4 and $L_G$=2 (All subsequent R-CPD/LAA-PMM employ these parameters by default), across varying numbers of P-users ($|U_P|$).
In the figure, (R-CPD with P-CPD) denotes that P-CPD is employed to construct P-Topo under the R-Topo produced by R-CPD; the other labels follow an analogous interpretation.

The delay of (R-CPD with P-CPD) / (R-CPD with DFCP) is lower than that of (LAA-PMM with P-CPD) / (LAA-PMM with DFCP). This is because R-CPD explicitly optimizes inter-satellite connectivity within the R-Topo.
As shown in Sec-\ref{sec_pcpd}-A, the benefits that R-CPD based R-Topo brings to P-Topo are greater than those brought by LAA-PMM based R-Topo.

The delay of P-CPD is always lower than that of DFCP, even the delay of (LAA-PMM with P-CPD) is lower than that of (R-CPD with DFCP). 
This indicates that even if the gain from R-Topo is poorer, the delay performance of P-CPD is still superior to DFCP. 
The advatage is attributed to the introduction of communication tendency $I_c$ in the weighted strategy of P-CPD for UG-Sats. 
The $I_c$ mechanism make UG-Sats evenly and frequently connecting to G-Sats, thereby achieving data downlink more efficiently.
Under R-CPD generated R-Topo, the to-GS delay of P-CPD is consistently within 2 time slots. 
Consequently, telemetry from any UG-Sat can be delivered to the GS within an average two time slots, which is critical for maintaining satellite operations.

\subsubsection{Satellite Ranging}
\begin{figure}[tbp] 
	\centering
	\includegraphics[width=0.82\linewidth]{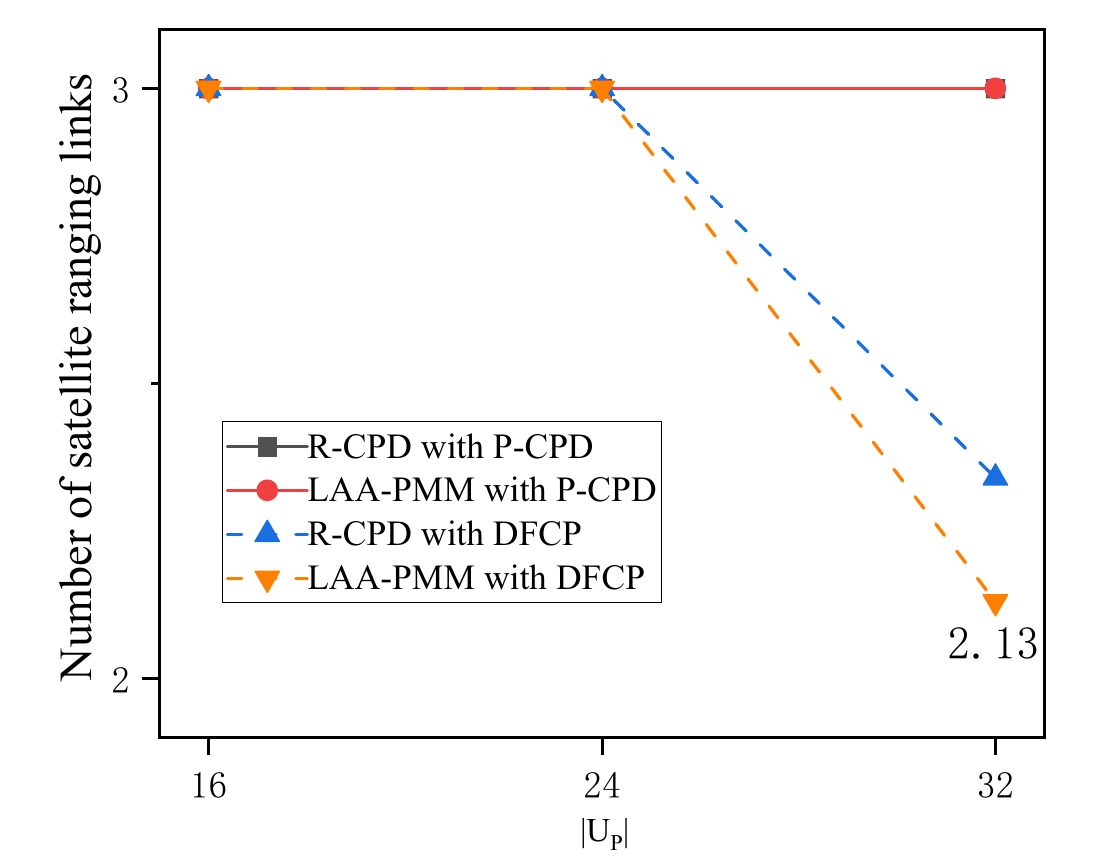}
	\caption{Number of satellites ranging links vs. $U_P$}
	\label{fig9}
\end{figure}
The number of ranging links and their geometric configuration jointly determine the orbit-determination accuracy of the constellation.
Fig.~\ref{fig9} reports the average number of ranging links per satellite. 
Under P-CPD, this value is exactly 3 in all cases. 
With 4 satellites in the scenario, 3 implies that each satellite performs ranging with every other visible satellite, thereby achieving the most diversified geometric ranging. 
Reference~\cite{8} has demonstrated that 3 ranging links per 5 minutes is sufficient to achieve meter-level positioning accuracy and converges within 23 days in this constellation.
In contrast, DFCP exhibits reduced ranging links when $|U_P|$=32. 
Consequently, P-CPD outperforms DFCP in terms of satellite ranging quality.

\subsubsection{P-User Communication}
\begin{figure}[tbp] 
	\centering
	\includegraphics[width=0.9\linewidth]{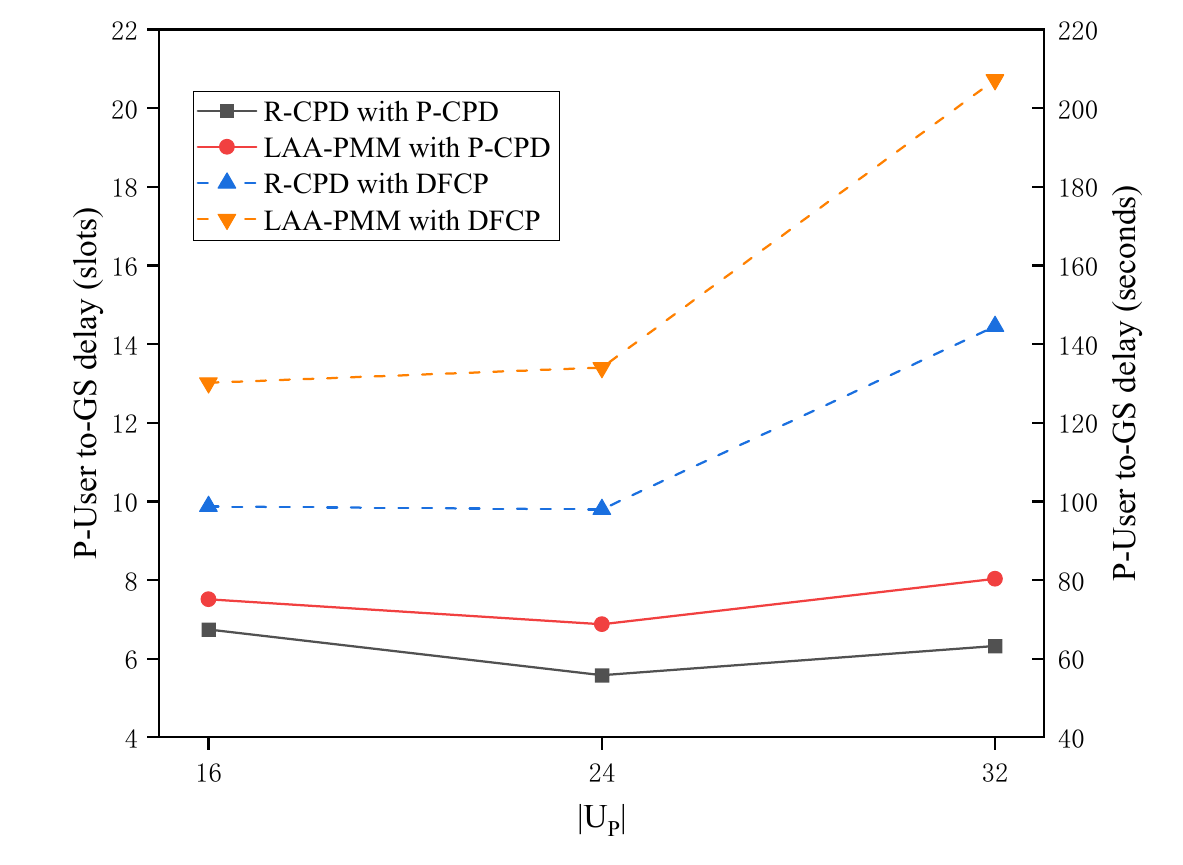}
	\caption{P-User to-GS delay vs. $U_P$}
	\label{fig10}
\end{figure}

Fig.~\ref{fig10} presents P-User average to-GS delay across varying $U_p$. 
Evidently, R-CPD yields substantially larger gains for P-Topo than LAA-PMM. 
The user delay under P-CPD is consistently lower than that under DFCP. 
This improvement arises for two reasons: 
first, the UG-Sats delay under P-CPD is already lower, so once a user access a satellite, the subsequent backhaul delay to the GS is naturally reduced;
second, the introduction of the user access tendency $I_u$ ensures that users associate with satellites uniformly over time, further suppressing to-GS delay.
Under (R-CPD with P-CPD), P-User data reaches GSs within 6-7 slots on average.

\subsubsection{P-User Ranging}
\begin{figure}[tbp] 
\centering
\includegraphics[width=0.9\linewidth]{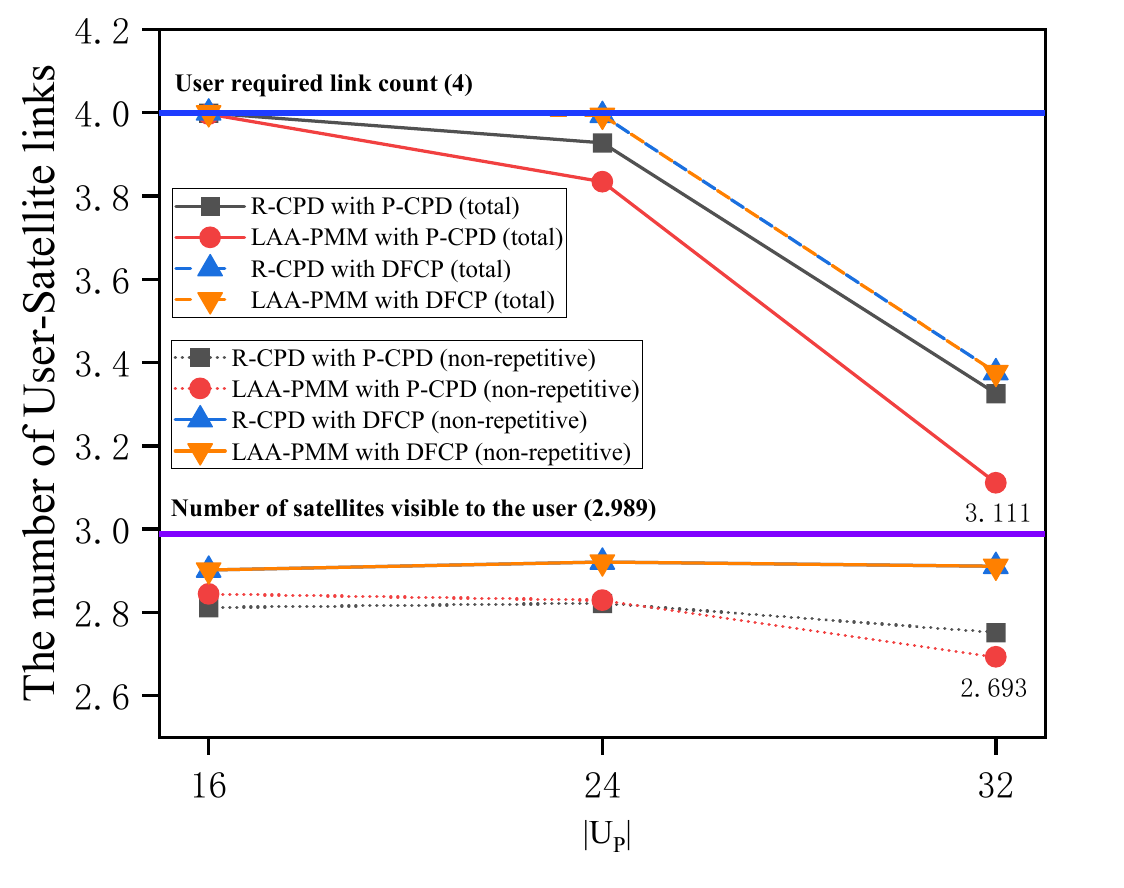}
\caption{Number of Satellite-User links vs. $U_P$}
\label{fig11}
\end{figure}
Navigation performance improves with both increased total PLs per user and more diverse user-satellite link geometries.
Fig.~\ref{fig11} reports the number of links provided to users across varying of $U_P$. 

In total numbers, both P-CPD and DFCP deliver nearly four links per user when $|U_P|$=16, which suffices to meet user demand in $U_P$.
At $|U_P|$=24, The number of user links in P-CPD exhibited a slight decrease. 
At $|U_P|$=30, link counts for both schemes decline. 

Considering geometry quality, each user is visible to 2.989 distinct satellites on average; 
link with all this distinct satellites yields the most diversified ranging geometry.
Although DFCP consistently maintains slightly more non-repetitive user-satellite links than P-CPD (by approximately 0.1–0.2), P-CPD still achieves nearly optimal geometric diversity.
Overall, DFCP marginally outperforms P-CPD in both total link counts and non-repetitive links numbers. 
This stems from DFCP’s pursuit of fairness among all nodes: as user population grows, users are given equal chances to connect, resulting in both higher total numbers and a preference for non-repeating links.
Nevertheless, DFCP falls significantly behind P-CPD in satellite delay and ranging quality. 
For an infrastructure, internal resources should be secured before better user service is considered, making P-CPD the more appropriate choice for the CLSI scenario.

\begin{figure}[tbp] 
	\centering
	\includegraphics[width=0.83\linewidth]{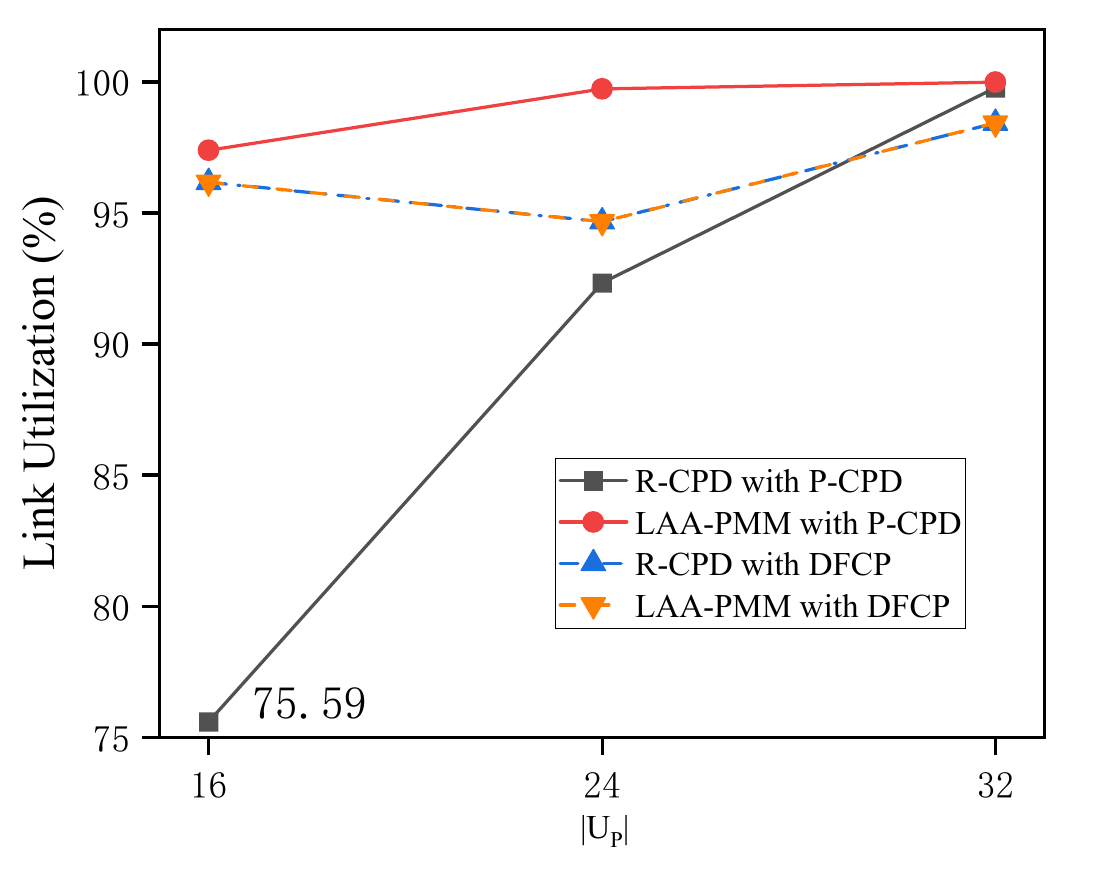}
	\caption{Link Utilization vs. $U_P$}
	\label{fig12}
\end{figure}

\subsubsection{Satellite Link Utilazation}
Fig.~\ref{fig12} plots link utilization under different $|U_P|$. 
Meeting performance requirements, lower link utilization directly translates to reduced energy consumption. 
At $|U_P| \leq$24, the (R-CPD with P-CPD) achieves the lowest link utilization. 
Compared to DFCP, it simultaneously provides significantly lower satellite and user delays while offering comparable quantity and diversity of user links.
This demonstrates (R-CPD with P-CPD)'s superior energy-saving capability under light P-User loads.
This efficiency stems from R-CPD's objective to maximize inter-satellite connectivity and P-CPD's weight-assignment strategy. P-CPD fully leverages the inherent benefits of R-Topo, eliminating the communication and ranging PLs that would otherwise be required without it.
Although both the (R-CPD with DFCP) and (LAA-PMM with DFCP) schemes achieve lower link utilization than (LAA-PMM with P-CPD), their satellite and user delays are significantly degraded. This indicates that their energy savings come at an unacceptable performance cost. 
Under heavy load ($|U_P|$=32), all schemes approach 100\% link utilization.
However, DFCP fails to balance internal satellite resource demands with user-serving resources. 
This results in substantial degradation in both satellite communication and ranging performance.
In contrast, P-CPD preserves the system's essential resources while still providing users with the best achievable service.

\subsubsection{Link Composition}
Table~\ref{Link Composition} details the link composition for the case $|U_P|$=32. 
It is evident that the link composition under (R-CPD with P-CPD) and (LAA-PMM with P-CPD) differ markedly.
Because the R-Topo generated by LAA-PMM is inferior to that of R-CPD, P-CPD operating on the LAA-PMM based R-Topo allocates additional PL(Sat,Sat)s to safeguard inter-satellite performance, thereby reducing PL(User,Sat)s. 
This demonstrates that P-CPD is sensitive to variations in R-Topo: it actively adapts its decisions to leverage the gains offered by the R-Topo.
Even in the extreme absence of R-Topo, P-CPD still reserves sufficient resources to maintain robust inter-satellite connectivity, confirming that P-CPD benefits from R-Topo yet remains fully operable without it. 
This also explains the significantly higher link utilization of (LAA-PMM with P-CPD) in Fig.~\ref{fig12}: P-CPD must allocate extensive resources to meet the satellite's internal requirements under LAA-PMM's poor R-Topo, while still attempting to maximize user service quality.
In contrast, DFCP exhibits an unchanged link composition, indicating insensitivity to R-Topo; it can only passively accept whatever benefits the R-Topo happens to provide.
\begin{table}[]
	\caption{Link Composition}
	\label{Link Composition}
	\centering
	\begin{tabular}{|c|c|c|}
		\hline 
		   & PL(Sat,Sat)s&  PL(User,Sat)s\\
		\hline
		R-CPD with P-CPD & 6.65 &106.45 \\
		\hline
		LAA-PMM with P-CPD & 10.22  &99.55 \\
		\hline
		R-CPD with DFCP & 5.04   &108.05 \\
		\hline
		LAA-PMM with DFCP & 5.04  &108.05 \\
		\hline
		
	\end{tabular}
\end{table}

\subsection{Further Discussion}

Based on the preceding performance analyses of R-Topo and P-Topo, we conclude the following:

\begin{enumerate}
	\item R-CPD is more suitable for this scenario than conventional communication CPD schemes such as LAA-PMM. The R-Topo produced by R-CPD returns the large-volume data from R-Users to the GS faster, demonstrating superior communication performance. In addition, the R-CPD based R-Topo provides greater benefits to P-Topo than LAA-PMM, improving the UG-Sats to-GS delay and other P-Topo metrics.
	\item P-CPD outperforms DFCP in CLSI scenarios. Under light P-User loads, (R-CPD with P-CPD) achieves superior satellite/user performance with lower resource utilization. As P-Users increase, DFCP provides slightly higher total/distinct user links at the cost of rapidly deteriorating satellite delay and ranging quality. Crucially, P-CPD preserves essential constellation resources while maintaining satisfactory user service levels.
	\item Owing to its specialized weight-assignment mechanism, P-CPD is sensitive to R-Topo and dynamically adapts its PL scheduling. It therefore fully exploits the advantages of R-Topo without becoming dependent on it, whereas DFCP remains insensitive to R-Topo and can only passively accept whatever benefits arise.
	\item R-CPD maximizes R-Topo's benefit conveyed to P-Topo, while P-CPD fully exploits this advantage;  this synergy demonstrates their tight coupling.
	\item The optimal values of the service, communication, and ranging constants in P-CPD ($C_u,C_c,C_r$) are scenario-dependent. As link schedules are computed on GSs days in advance and uploaded to satellites, this offline process imposes no computational constraints. Thus, scenario-specific constants can be optimized via simulated annealing, AI-based search, or other techniques without computational limitations.
\end{enumerate}

\section{CONCLUSIONS}
\label{sec_conclusion}

Satellites can be deployed at Earth-Moon three-body orbits to construct a Cislunar Space Infrastructure (CLSI). 
In this paper, we study the CPD problem in CLSI.
Since satellites are equipped with reflector and phased array terminals, we discuss the CPD for the R-Topo and P-Topo separately. 
We employ R-CPD using ILP model to generate an superior R-Topo that satisfies the communication needs of reflector users. 
Based on the R-Topo, we implement a P-CPD using maximum weight matching to meet satellites’ ranging and communication needs while providing communication and navigation services to phased array users. 
Performance analysis reveals a strong coupling between R-CPD and P-CPD. 
By operating in concert, R-CPD and P-CPD deliver high-quality services to users while simultaneously meeting the satellite’s internal resource requirements.
The proposed CPD scheme provide strong technical support for constructing CLSI.

However, this paper focuses solely on scheduling links for satellites within cislunar space at three-body orbits. Integrating this cislunar constellation with potential future interplanetary deep-space constellations will be a promising direction for future work.

\bibliographystyle{IEEEtran}
\bibliography{references}

\end{document}